\documentclass{article}
\usepackage{graphicx} 
\usepackage{graphicx}
\providecommand{\keywords}[1]{\textbf{Keywords:} #1}
\usepackage{amsmath}
\usepackage{amssymb}
\usepackage[OT4,T1]{fontenc}
\usepackage[utf8]{inputenc}

\usepackage{authblk}
 
\begin{document}

\title{Why I don't like logistic equation?}
\author{Janusz Uchmański

Cardinal Stefan Wyszyński University

Wóycickiego 1/3, 01-938 Warsaw, Poland

j.uchmanski@uksw.edu.pl}

\maketitle

\begin{abstract}
 The logistic equation treats the definition of its variable freely. It is undoubtedly the population density. The logistic equation\providecommand{\keywords}[1]{\textbf{Keywords:} #1} in the differential version requires continuity of processes occurring in the population. In the differential version, we do not know what happens to the population between time steps. When filling these gaps, models with features different from the logistic equation appear. If we have a micro-scale model describing the fate of individuals and interactions between them, then the macro description for this model is not the equation for population density (e.g. the logistic equation) but the model of matter circulation or energy flow through this system. We lack mathematics to describe purely biological processes.
\end{abstract}


\section{Introduction}
\vskip 0.3 cm
In the differential version, the logistic equation proposed as a description of population dynamics has the following form:
\begin{equation}\label{1}
\frac{dN}{dt}=r(1 - \frac{N}{K})N.
\end{equation} 
In the difference version, the logistic equation can have two forms. In the case of describing a population with overlapping generations, it will be the following equation:
\begin{equation}\label{2}
N_{t+1}=N_t+r(1-\frac{N_t}{K})N_t.
\end{equation}
Equation (2) is the difference equivalent of the logistic equation in the differential version, because the latter does not say anything specific about the division into generations. However, if these are non-overlapping generations, we will  get a slightly different form:
\begin{equation}\label{3}
N_{t+1}=r(1-\frac{N_t}{K})N_t.
\end{equation}
\section{What does the logistic equation describe?}
\vskip 0.3 cm
It is worth noting at the outset that population dynamics is a problem that has the status of "significant" in ecology. In physics, the number of gas particles is not a significant problem. At a fixed temperature, it will always be the same in a unit of volume. In ecology, the number of individuals in a population changes because individuals arise in different ways and die. Both of these events are obligatory for each individual. In addition, individuals produce offspring individuals, and although some individuals may not succeed, this is an event that each individual must deal with. Gas particles simply persist.

What is the variable $N$ that appears in these equations as a state variable? There is a lot of confusion in the ecological literature on this issue. Ecologists and especially mathematicians treat this problem very loosely and most often say at the same time that it is the population density and the population number, without considering whether this does not make a difference. What is more, the term population size is encountered in the literature without any explanation of what it means, sometimes you can even find a phrase that sounds something like this: the dynamics of population $N$ is described by equation \eqref{1}.
	
It must be clearly stated that $N$ is the population density. There is no doubt about this in the case of equation \eqref {1}. The solution of this differential equation is a continuous real function $N$($t$). The population number would be a natural number, so the logistic equation in the differential version describes changes in population density. The situation is a bit more complicated in the case of the difference version of the logistic equation. Let us assume that at a certain time step $N_t$ is a natural number. Then sooner or later in one of the next time steps this variable will become a real number. This will be the result of multiplying by $r$ and the fractional value of the quotient $N_t$/$K$. However, one can argue that after all, the value of the variable $N_{t+1}$ obtained from equation \eqref{2} or \eqref {3} can always be rounded up or down. Consider the following example. Let 100 individuals live on 75 $m^{2}$ of land. Then the density will be 1.3333… individuals per $m^{2}$. We round this to 1 individuals per $m^{2}$. Now let's multiply this density by 75 $m^{2}$. We will get 75 individuals in the entire population. How did 25 individuals disappear from the population and why. What's more, we can ask which individuals disappeared - males, females, young, old, etc. So it's better to agree that the logistic equation in each version describes changes in population density. Here, for the first time, we encounter the problem of individuality of elements belonging to the biological population. Each biological individual has its own individual history behind it, and also its own, unique further life ahead of it. The question arises at what level of development of the biological organization of the organism do we encounter the need to take into account its individuality in the mathematical description.

Ecologists rarely know the number of individuals in a population. It is relatively easy to measure population density, and it is almost always a measurement of the local value of this variable. Models such as the logistic equation assume the existence of certain laws governing changes in density. However, density essentially requires prior knowledge of the number of individuals in the population. It seems, therefore, that population number is a primitive concept in relation to its density, and calculating density requires one additional division operation. Let us also note that in order for density to carry ecologically significant information, it is necessary to assume that the population is well-mixed and that there are no phenomena and processes occurring on a local scale.

\section{The continuity problem}
\vskip 0.3 cm
All processes described by the logistic equation in the differential version occur continuously in time. The solution to this equation is also a continuous function. Let's look for a real population that would satisfy the continuity postulate. Let's imagine a culture of bacteria. On an ecological scale, this is a macroscopic object and if the culture is large enough, we can safely assume that at every moment something is happening to one of the microscopic components of this system, some cell is dividing and another one dies. And other populations. For example, the human population. It is so numerous that we can also assume that at every moment someone is born or someone dies somewhere. |What's more, we can try to calculate the global density of this population by dividing the number of people on Earth, which we know more or less exactly, by the area of land inhabited by people. But what significance does it have for a woman from Irkutsk who wants to give birth to a child that one of the residents died in a nursing home in Warsaw?

It is clear that density must meet another condition. It must be the density of such a group of individuals between whom real interactions take place, either directly or through the exploitation of common resources. This can be achieved by narrowing the size of the space occupied by the population, but then there is a huge chance that our population will cease to meet the condition of continuity. The problem of continuity of events in the population can also be viewed from the point of view of the individual. An individual dies only once and no individual reproduces continuously.

A biologically justified approximation of continuity could be sought in populations with overlapping generations. Island populations would be particularly interesting here, as this would solve the problem of calculating actual population density. The number of co-occurring generations in long-lived species can be quite large, as, for example, in the case of the mew gull \textit{Larus canus} (\cite{bukacinski1}, \cite{bukacinski2}), which inhabits islands on the Vistula River. However, the continuity condition is violated here by the existence of only one brood per season (sometimes repeated only once) and the synchronization of the life cycles of different generations, correlated with environmental seasonality. In turn, in the population of the bank vole \textit{Clethrionomys glareolu}s inhabiting the Crabapple Island in the Masurian Lakeland in Poland (\cite{petrusewicz}, \cite{bujalska1}, \cite{bujalska2}), offspring production by individuals from several cohorts occurs throughout the summer season from April to September. However, in both cases, the numbers of individuals in all categories are relatively small, which is a consequence of the dispersal nature of these populations.

Perhaps the only system that satisfies the condition of continuous ecological processes is a colony of social insects. A eusocial system of organization greatly simplifies the relationships between individuals. Because worker bees do not reproduce, it is difficult to speak of competition between them. They dedicate themselves to working cooperatively for the benefit of the entire colony, which largely makes it independent of environmental influences. The queen bee continuously produces eggs, with an intensity that varies depending on the season. In summer, these can reach hundreds of eggs per day, while in winter, only a few dozen. The short lifespan of workers - two or three weeks in summer and a few months in winter - along with the queen's intense efforts, results in a nearly continuous process of worker waxing and waning, and this category itself consists of many overlapping and relatively short-lived generations. Of course, many other details of the biology of this group of insects remain aside, which become important if we attempt a description on a long time scale, for example the death of the queen, the formation of daughter colonies, or the need to produce drones.

In the logistic equation in the difference version, all processes proceed discontinuously. In equation \eqref{2}, we transfer the density from the previous time step to the next time step by adding something that depends on the density in the previous time step. In equation \eqref{3}, the density in the next time step is a certain part of the density in the previous time step. In both cases, the question arises what happens to the population between adjacent time steps. Does the population exist or disappear, or do the individuals of the population hibernate? After all, we are describing something that exists continuously. Do important events in the population occur only at certain moments, and do ecological processes die out between them? In short, the logistic equation in the difference version provokes us to fill these temporal gaps with biological content (\cite{kalmykow1}, \cite{kalmykow2}).

Another doubt arises. Let us imagine that the density of the actual population increases, reaches a maximum, and then falls. We mark the points corresponding to the successive time steps. Let the same densities occur on both sides of the maximum in certain time steps. How will the logistic equation in the difference version know that it is in the phase of population growth at one time and decline at another, since in both cases we have the same density. One can argue that oscillations of this nature are not among the solutions of the logistic equation in the difference version, but such oscillations not caused by the presence of a predator are in the repertoire of the dynamics of actual single populations. This shows that the choice of one modeling method and not another limits our field of vision of the problem. Unfortunately, every choice of mathematical description method is burdened with this imperfection, but the logistic equation is exceptionally harmful here. After all, it is the logistic equation that was chosen to justify our ideas about population dynamics. The solution of the logistic equation is a model of what ecologists \textit{a priori} understand by the concept of population regulation. This is probably because of its simplicity rather than realism (\cite{winley}), even though it was noticed more than 80 years ago that the logistic equation fits the existing empirical data poorly (\cite{feller}). It also turns out that the mathematical simplicity of the logistic equation is deceptive. For example, it is necessary to make very biologically unrealistic assumptions if we want to create a stochastic version of the logistic equation (\cite{winley}).

Perhaps it would be much more honest if we proceeded differently. Let us abandon the explicit, mathematical formulation of the law governing changes in population density. Instead, let us focus on the mathematical description of the lives of individuals, taking into account their interactions, and only occasionally count how many of them we have in the population. If our description includes the births and deaths of individuals, then it will be less burdened by all the limitations to which the logistic equation is subject.

\section{Gap filling results}
\vskip 0.3 cm
Nature is very diverse. That is why we have just as many options when we start filling the gaps in the logistic equation. We can go to a very high level of detail and make a model of a very specific situation involving a specific species, or we can limit ourselves to certain essential features of the system, i.e. those that distinguish it from a physical system. As for modeling methods, there is not much choice. It will be a mixture of analytical methods and computer simulations in various proportions. However, there is little doubt that if we assume that the population consists of points moving randomly in space, and the meeting of two of them results in the appearance of a new point or the disappearance of one of them, then using quite complicated argumentation it can be shown that the dynamics of such a system will be described by the logistic equation (\cite{nedorezov}, \cite{lachowicz}, \cite{bustos}). Such a set of points is far from the real population.

Let's take a population of insects with non-overlapping generations (\cite{uchmanski2}). In spring, eggs develop into larvae, which grow using resources with well-defined dynamics. We introduce competition between larvae. As a result, we get an unequal partitioning of resources among competing individuals according to such principles as to obtain well-documented in ecological literature features of the distributions of competitors' weights (\cite{uchmanski1}). In autumn, larvae go through the pupal stage, which transform into adults. The latter do nothing else but lay eggs, and from them the next spring the next generation of larvae will arise. In such a model, we do not use any equations for  direct description of  changes in the population number. We simply count individuals in the population from time to time.

The range of solutions of such a model is much richer than in the logistic equation, and in addition we can show what is their cause (\cite{uchmanski3}). Among the solutions there is a solution resembling the classic solution of the logistic equation, which is recognized by ecologists as a pattern of population regulation, but there are also differently expressed oscillations of numbers, as well as population extinction. The type of solution depends, for example, on the dynamics of the resources, which in turn can be linked to the features of the real environment, for example with the features of a seasonally changing environment or a seasonless environment. Individual variability becomes very important. A population with strongly expressed individual variability behaves completely differently than a population composed of identical individuals. The difference between local and global competition is visible.

Filling the gaps that exist in the difference scheme of the logistic equation takes us to a completely different world, the traces of which are not visible among the solutions of the logistic equation. The decision as to what is closer to ecological reality is probably obvious. 

There are, however, costs to such a procedure. Mathematics provides us with wonderful tools for analyzing the properties of differential and difference equations. Filling the gaps, we must necessarily resort to inelegant computer simulations, because such is the biological essence of the described system, the essential feature of which is that it consists of individuals. Simulations also require providing numerical values of the model parameters. Such data exist in the ecological literature, but they are usually sufficient only to create a model of a very specific, individual situation. If we want to build a model of a more general nature, we face a serious problem, although we can refer to the analysis of its properties in the parameter space. The question arises about the importance of the results obtained from computer simulations. Is a computer simulation equivalent to an experiment with living nature? One may have doubts about this. Experimentation allows us not to worry about many elements of the studied system, nature itself provides them to us.

\section{The problem of scale}
\vskip 0.3 cm
I believe that my experience with attempts with filling the gaps in the logistic equation entitles me to state that the logistic equation is not an approximation of ecological reality. However, we have the right to expect in the natural sciences the agreement of descriptions of reality made at different scales. An example of this in physics is the agreement of descriptions of gas properties provided on the one hand by statistical physics and on the other by thermodynamics. There is also room for this in ecology. However, it is a mistake to assume that the logistic equation is a description on a macro scale of what results on a micro scale from filling the gaps in the logistic equation. The logistic equation is not a description on a macro scale. Both approaches try to describe the same. There is no secret behind the transition from population number to population density; space does not participate in it in a significant way. Density is a simple division of population number by the measure of space. If necessary, space can participate in a significant way in the description of interactions between individuals. A model of an ecological system on a macro scale is a model of the circulation of matter in ecological systems or the flow of energy through them. And since in the latter we describe the result of the collective activity of a great many organisms, in this case we can enjoy the continuity of processes and the possibility of using differential equations. Let us note that, for example, the problem of the continuity of the course of ecological processes disappears if we move from the description of the dynamics of the population number or density to the description of the dynamics of the total biomass of all individuals in the population, because the biomass of an individual changes continuously to a good approximation, while it itself comes into being, reproduces and dies only at certain moments.

\section{Conclusion}
\vskip 0.3 cm
The text presented here is a kind of cry for help. Until I realized everything I have written here, I had no problem understanding the models of classical theoretical ecology. Now I get lost in the interpretations of these models. At such moments I feel like dropping everything, going to the forest and looking at living nature.

In 1790, in his "Critique of the Powers of Judgment", Immanuel Kant said that Newton for a blade of grass had not yet been born. Almost 250 years have passed and nothing has changed. What's more, in the middle of this period mathematicians came up with the idea of using the same mathematical methods to describe the dynamics of ecological systems that were created for the needs of physics. The effects of this are still felt today, as exemplified by the logistic equation and all models of classical mathematical ecology. However, it remains true that no mathematics has yet been created that would meet the requirements of biology, and ecology in particular. Of course, in biology there is a lot of room for physical phenomena (for example in the physiology of an organism), but the essence of ecology, which is that ecological systems consist of individuals whose features are the product of natural selection and which, on their own account, although in interactions with other individuals, implement their algorithm called life, remains beyond the reach of mathematics.

\vskip 0.3 cm

\end{document}